\documentclass{emulateapj}
\usepackage{graphicx,amsmath,amsfonts,amssymb,subfigure,natbib,epsfig}

\newcommand{\avgfeh}{$\langle$[Fe/H]$\rangle \ $}
\newcommand{\feh}{[Fe/H] } 
\newcommand{\sigmafeh}{$\sigma$([Fe/H]) \ }
\def\Sec{${}^{\prime\prime}$\llap{.}}

\newcommand{\cat}{\ion{Ca}{2}\ }

\begin{document}

\title{Metallicity Evolution of the Six Most Luminous M31 Dwarf Satellites}

\author{Nhung\ Ho\altaffilmark{1}}
\author{Marla\ Geha\altaffilmark{1}}
\author{Erik~J.\ Tollerud\altaffilmark{1,3}}
\author{Robert Zinn\altaffilmark{1}}
\author{Puragra\ Guhathakurta\altaffilmark{2}}
\author{Luis~C. Vargas\altaffilmark{1}}

\altaffiltext{1}{Astronomy
  Department, Yale University, New Haven, CT~06520.
  ngocnhung.ho@yale.edu, marla.geha@yale.edu}

\altaffiltext{2}{UCO/Lick Observatory, University of California,
   Santa Cruz, 1156 High Street, Santa Cruz, CA~95064.}
   
\altaffiltext{3}{Hubble Fellow}

\begin{abstract}
\renewcommand{\thefootnote}{\fnsymbol{footnote}}

We present global metallicity properties, metallicity distribution
functions (MDFs) and radial metallicity profiles for the six most
luminous M31 dwarf galaxy satellites: M32, NGC~205, NGC~185,
NGC~147, Andromeda~VII, and Andromeda~II.  The results presented are
the first spectroscopic MDFs for dwarf systems surrounding a host galaxy other than
the Milky Way.  Our sample consists of individual metallicity
measurements for 1243 red giant branch (RGB) member stars spread
across these six systems.  We determine metallicities based on the
strength of the \cat triplet lines using the empirical calibration of
Carrera et al.~(2013) which is calibrated over the metallicity range
$-4 < $[Fe/H]$<+0.5$.  We find that these M31 satellites lie on the
same luminosity-metallicity relationship as the Milky Way dwarf
satellites.   We do not find a trend between the internal metallicity
spread and galaxy luminosity, contrary to previous studies.    The
MDF widths of And~II and And~VII are similar to the MW dwarfs of
comparable luminosity, however, our four brightest M31 dwarf are more
luminous than any of the MW dwarf spheroidals and have broader MDFs.
The MDFs of our six M31 dwarfs are consistent
with the leaky box model of chemical evolution, although our
metallicity errors allow a wide range of evolution models.  We find a
significant radial gradient in metallicity in only two of our six
systems, NGC 185 and Andromeda II, and flat radial metallicity
gradients in the rest of our sample with no observed correlation
between rotational support and radial metallicity gradients.    While
the average properties and radial trends of the M31 dwarf galaxies
agree with MW counterparts at similar luminosity, the detailed MDFs
are different, particularly at the metal-rich end.

\end{abstract}


\section{Introduction}

The increasing number of spectroscopic observations of individual stars available in nearby Milky Way dwarf galaxies allows for the detailed characterization of both the kinematic and metallicity distributions in these low luminosity systems \citep{Tolstoy2004, Helmi2006,walker2007, Simon2007a, Battaglia2011, Kirby2011,Simon2011}.  While kinematics provides a snapshot of the current dynamical state of a galaxy, the evolutionary history of a galaxy is encoded in its metallicity.   More specifically, the distribution of metallicity within a galaxy hold clues to its star formation, gas accretion, and gas expulsion history \citep[e.g.][]{Prantzos1998,Matteucci2001,Lanfranchi2004}

Deep photometric observations and chemical abundance studies of individual stars in many Milky Way (MW) satellites have allowed for the detailed study of the formation histories of these systems \citep[e.g.,][]{Aaronson1985,Buonanno1985,Smecker1994,Stetson1998}.  For example, in two of the more massive MW dwarfs, Sculptor and Fornax, detailed chemical abundances have been determined for a significant portion of the red giant branch stars in both galaxies \citep{Tolstoy2004,Pont2004,Gullieuszik2007,Battaglia2008,Kirby2009,Starkenburg2010}.  Combining these abundances with deep photometric observations of their stellar populations allows for the deduction of their complete star formation histories (SFHs) \citep{Deboer2012,Coleman2004,Deboer2012b}.  However, detailed photometric and spectroscopic work are time intensive and only recently have the wider MW dwarf population been observed in such detail \citep{Dolphin2005, Holtzman2006,Brown2012,Gilmore2013}.  

The shapes of spectroscopic metallicity distribution functions for Milky Way dwarf galaxies, along with age estimates of their constituent stellar populations, demonstrate that these objects have a wide range of star formation histories \citep{Gallart1999,Aparicio2001,Tolstoy2004, Helmi2006, Battaglia2011}.  Despite variations in individual star formation histories, the Milky Way dwarf galaxy satellites show a tight, linear luminosity-metallicity relation over more than three orders of magnitude in luminosity \citep{Caldwell1992,Kirby2008b, Mcconnachie2012}.  The trend between galaxy luminosity and galaxy metallicity also extends to more massive galaxies, with a flattening for the most massive objects \citep{Skillman1989,Tremonti2004, Andrews2013}.   Comparison to chemical evolution models further suggest that significant gas outflows are needed to explain the metallicities of most, though not all, of the Milky Way dwarf galaxies \citep[e.g.,][]{Tolstoy2001,Winnick2003,Koch2006,Kirby2011}.  

Dwarf galaxies around the Andromeda (M31) galaxy are sufficiently nearby to resolve individual stars, providing a second satellite system in which MDFs can be determined.  M31 is host to a larger number of luminous satellites as compared to the Milky Way (13 M31 satellites versus 7 Milky Way satellites brighter than $M_V = -10$), likely due to the larger total galaxy mass of the M31 system \citep{Watkins10a,Yniguez2013}.  The M31 satellites appear to follow the same relationship between luminosity and average metallicity seen for the MW dwarf galaxies from both photometric and spectroscopic metallicity measurements \citep{Caldwell1992,Kalirai2010, Mcconnachie2012,Collins2013}.  While several studies have presented average metallicities \citep{Kalirai2010, Collins2013}, and in some cases binned radial metallicity profiles \citep{Geha2010}, none determined spectroscopic MDFs for M31 satellites.

Characterizing the MDF of a galaxy requires individual metallicity measurements for a large sample of individual stars.  High resolution spectroscopy, the gold standard for determining stellar chemical abundances, is far too expensive to build up significant metallicity samples in systems beyond a few tens of kpc \citep[e.g.,][]{Shetrone2003, Frebel2010}.    The first large, homogeneous metallicity determinations using purely spectroscopic indicators were of globular clusters in the Milky Way by \citet{Rutledge1997b} using an empirical calibration between the strength of the Ca II triplet (CaT) line strength and [Fe/H], a method pioneered by \citet{AD1991} and \citet{Olszewski1991}.

This empirical calibration fails at the low metallicity end where the \cat lines are much narrower and non-LTE effects begin to dominate in the stellar photosphere \citep{Gray2005}. \citet{Starkenburg2010} updated the CaT calibration, allowing for a non-linear relationship between the CaT equivalent widths, over the metallicity range $-4.0 <$ [Fe/H] $< -0.5$, basing the calibration on synthetic stellar spectra.  However, this still does not fully cover the observed range of metallicities observed in the dwarf satellites and relies on synthetic spectra, which cannot fully capture the complex physics involved.  \citet[][hereafter C13]{Carrera2013} has recently expanded the empirical calibration of the CaT by combining calibration data from metal-rich open clusters and extremely metal-poor halo stars to cover the range $-4.0 <$ [Fe/H] $< +0.5$.

Here we present the metallicity distributions functions (MDFs) based on \citet{Carrera2013} calibrated \cat triplet lines for the six brightest dwarf galaxy satellites around M31.  These are the first MDFs presented for any satellite of M31.   The data were taken homogeneously, using the Keck/DEIMOS spectrograph, and have been previously used to study the kinematics in these systems: M32 by \citet{Howley2013}, NGC~205 by \citet{Geha2006a}, NGC~147 and NGC~185 by \citet{Geha2010}, Andromeda~II (And~II) by \citet{Kalirai2010,Ho2012} and Andromeda~VII (And~VII) by \citet{Tollerud2012}.   M32, And~II, and And~VII are all part of the Spectroscopic and Photometric Landscape of Andromeda's Stellar Halo (SPLASH) Survey.  

The paper is organized as follows.  We describe the spectroscopic datasets and membership selection process in \S\,\ref{methods}.  In \S\,\ref{sec_calibration}, we describe in detail the CaT calibration used to determine [Fe/H] for stars in our sample.   We next discuss the resulting metallicity distributions for M31 satellites including average global metallicities (\S\,\ref{ssec_global_feh}), global metallicity dispersions (\S\,\ref{ssec_global_dispersion}), the MDFs and a comparison to a simple chemical evolution model (\S\,\ref{ssec_mdfs}), and radial metallicity profiles (\S\,\ref{ssec_radial}).

\section{Spectroscopic Observations and Membership Selection} \label{methods}

We present individual stellar metallicities for the six brightest M31 dwarf galaxy satellites based on the strength of the calcium triplet lines near  $8550\mbox{\,\AA}$.  Our data were taken using the Keck~II 10-m telescope and the DEIMOS spectrograph \citep{faber2003a} with the 1200~line~mm$^{-1}$\,grating covering a wavelength region $6400-9100\mbox{\,\AA}$.  The spectral dispersion was $0.33\mbox{\,\AA}$ pixel$^{-1}$, equivalent to R=6000 for our $0$\Sec$7$ wide slitlets, or a FWHM of $1.37\mbox{\,\AA}$. The spatial scale was $0$\Sec$12$~per pixel.  We refer the reader to \citet{Geha2010, Tollerud2012} and \citet{Howley2013} for details on the data reduction.

All spectra from the six M31 dwarf galaxies in our sample are sourced from previously published work with the partial exception of And~VII.  In all cases, the published work focused on kinematics derived from the Keck/DEIMOS spectra rather than metallicities.  We summarize relevant targeting and membership details below and describe our new And~VII data in \S\,\ref{ssec_and7}.  

\begin{figure}[!t]
\centering
\includegraphics[width=.48\textwidth]{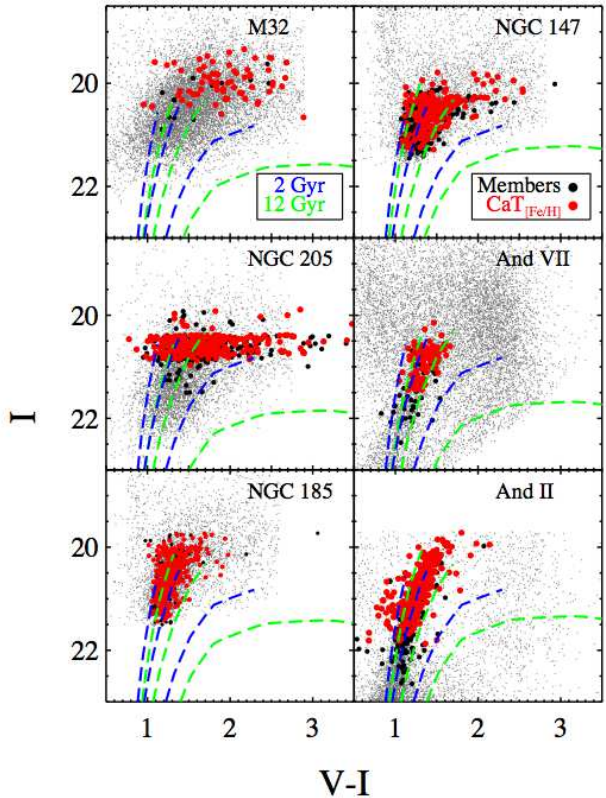}
\caption{Color Magnitude diagram for the six most luminous M31 dwarf galaxy satellites showing all photometric data (grey circles), member stars (filled, black circles), and member stars with measured calcium triplet metallicity (filled, red circles).  Overlaid are 2 Gyr (blue, dashed line) and 12 Gyr (green, dashed line) Padova isochrones \citep{Girardi2002} with \feh = $-$2.0, \feh = $-$1.0, and \feh = 0.0, respectively.} \label{CMD_fig}
\end{figure}

Target selection for NGC~205, NGC~147 and NGC~185 were based on $R$- and $I$-band photometry from the CFHT12K imager.  Target selection for And~II and And~VII were based on imaging from the Kitt Peak Mosaic camera in the Washington System \textit{M} and \textit{T$_2$} filter, as well as the \textit{DDO51} intermediate-band filters centered near the surface gravity dependent Mgb and MgH absorption lines, which allows us to separate foreground MW dwarf stars and target M31 giant stars \citep{Ostheimer2003}.  Target selection was based on a star's position on the Color Magnitude Diagram (CMD) relative to a metal-poor 13 Gyr isochrone \citep{Girardi2002} and position on the \textit{M$-$DDO51} and \textit{M$-$T2} color-color diagram \citep{Tollerud2012}.  As seen in Figure~\ref{CMD_fig}, targets cover a good fraction of the RGB.  

We have used the memberships as determined from each of the galaxy source papers:   for M32 in \citet{Howley2013}, NGC~205 in \citet{Geha2006a}, NGC~147 and NGC~185 in \citet{Geha2010}, And II in \citet{Ho2012}, and \citet{Tollerud2012} plus \S\,\ref{ssec_and7} for And~VII.    Five of our galaxies: NGC~205, NGC~147, NGC~185, And II, and And VII each contain over $>$100 member stars while M32 contains over 60 member stars, all have spectra with an average per pixel S/N of 6.0. The general method used for membership determination in these dwarfs relies on combined criteria of velocity, position in CMD space and gravity sensitive line indicators.  While most of our dwarf galaxies have radial velocities which placed them outside of the stellar velocity peaks of M31 and the MW, they still overlap with the wings of both distributions.  In order to establish membership, a combination of three criteria were used: line of sight velocity to establish membership within the dwarf, strength of the Na I absorption line at $\lambda8190 \mbox{\,\AA} \ $ to remove foreground dwarf stars from the sample, and the distance to a fiducial isochrone to remove additional contamination.  Details on final sample used in our spectroscopic metallicity analysis can be found in \S\,\ref{metal_samp}.

\subsection{Keck/DEIMOS Observations for And VII}\label{ssec_and7}

\citet{Tollerud2012} presented 136 members for And~VII based on two Keck/DEIMOS masks.   We supplement these data with an additional two DEIMOS masks observed between September 15--17, 2012.  The target selection and data reduction, as well as the probabilistic membership determinations, mirror that of \citet{Tollerud2012}.  Overall, we identify a total of 70 new RGB members in And VII, a significant increase to the previous sample.  Our final And VII sample consists of 206 member stars.

\begin{figure}[!h]
\centering
\includegraphics[width=.35\textwidth,angle=90]{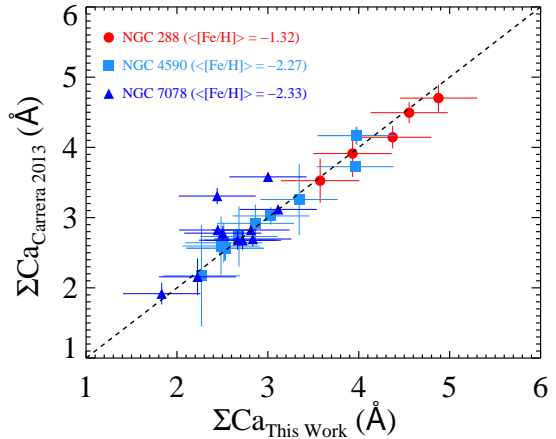}
\caption{Comparison between the equivalent width calculation for the CaT lines at 8542 $\mbox{\,\AA}$ plus 8662 $\mbox{\,\AA}$ using our transformed Gaussian profile fits and that of the globular cluster sample in \citet{Carrera2013} for three overlapping globular clusters.  The dashed, black line corresponds to a 1:1 relation between our measurements and that of C13.  Despite differences in continuum, bandpass, and functional form used to fit the EWs of the CaT lines, we are measuring the same quantities as C13.} \label{COMPARISON_fig}
\end{figure}

\section{Calcium Triplet Metallicity Calibration}\label{sec_calibration}

To determine the metallicity ([Fe/H]) for individual stars across our sample, we utilize the near-infrared calcium CaT lines, which are an effective proxy for measuring the \feh of RGB stars across a wide span of stellar ages \citep{Pont2004,Cole2004, Carrera2007,Saviane2012}.  We measure equivalent widths for the CaT lines and utilize the C13 metallicity calibration to convert observed CaT equivalent widths and I-band luminosity of RGB stars to \feh.  

To measure the equivalent width of the CaT lines, previous works have either directly integrated the continuum-subtracted flux, or have fit the CaT lines with an analytic function and then integrated over the area of the resulting fit.  A typical analytic function used to fit these lines is the Gaussian; however, as noted by previous authors \citep{Rutledge1997b,Cole2004,Carrera2007}, the Gaussian profile is a poor fit to the CaT line shapes as metallicity increases due to the broadening at the wings of the line.  To correctly measure the line shape, \citet{Cole2004,Carrera2007} and \citet{Saviane2012} showed that a sum of a Gaussian plus Lorentzian (G+L) provides a much better fit at both the low metallicity, where line profiles have a more Gaussian shape, and at the high metallicity, where the profiles are more Lorentzian.

While the G+L function provides the best approximation to the shape of the true line profile for high S/N data (S/N$\ge$25/pixel),  stars within our sample have lower typical S/N ($< 10$/pixel).   The G+L profile is less stable at low S/N as the effects of imperfect sky subtraction creates artificial depressions, or enhancements, in the outer wings of the third Ca line at 8662$\mbox{\,\AA}$.  This effect is apparent when comparing the trend between EW$_{8542}$ to EW$_{8662}$, where we did not observe a linear relation between the two lines as in C13.  To mitigate this effect, we limit our fit to a pure Gaussian, which has a sharper boundary at the wings and thus, is more robust against instrumental wing dampening due to the presence of errant pixels.   

Despite using a Gaussian to fit the line profiles, we still recover the true EWs of the lines by utilizing a linear relation between the EWs measured using a pure Gaussian and that using a G+L profile \citep{Cole2004, Carrera2007, Saviane2012}.  To determine this relation, we used a sample of 22 globular clusters observed with DEIMOS from \citet{Kirby2008a}, with a combined 429 RGB members and a minimum S/N = 25 per pixel, we calculated the EWs for each of the two CaT lines using the same continuum and line definitions from \citet{AD1991} for both the Gaussian and G+L profiles.  We find, by imposing passage through the origin (0,0), that EW$_{G+L} = 1.114(\pm0.01)\times {\rm EW}_{G}$, which is similar to the relation found by \citet{Saviane2012}.  Using this relation, we transformed our Gaussian EW measurements into G+L EW measurements.  

To ensure that our EW measurements are correct with the inclusion of the correction factor from Gaussian to G+L, we compare our line measurements against that of C13, who used a G+L for their profile measurements, for three overlapping globular clusters: NGC 288, NGC 4590, and NGC 7078 with a total 29 overlapping RGBs stars.   As shown in Figure \ref{COMPARISON_fig},  we find that our measurements, with the correction factor applied, are consistent with that measured by C13.  Verifying that, despite the differences in continuum and line bandpass definitions, we are measuring the same quantities as C13, we can now apply their calibration to our M31 dwarf RGB sample.

We use the line and continuum regions defined by \citet{AD1991} for the CaT lines at 8542 $\mbox{\,\AA}$ and 8662 $\mbox{\,\AA}$.  While the bandpass region used in our work is different from that of C13, we determined that this does not affect the recovered metallicity of the sample stars.  We exclude the weakest line ($\lambda = 8498\mbox{\,\AA}$) from this work due to the aforementioned low S/N of our data; including the weakest CaT line introduces more noise than signal into the total EW.  Adopting the C13 method, we determine the unweighted sum of the two lines as the total equivalent width 

\begin{eqnarray}
\Sigma \rm{Ca} = {\rm EW}_{8542}+{\rm EW}_{8662}.
\end{eqnarray}

In addition to the combined equivalent width, $\Sigma \rm{Ca}$, conversion to \feh also depends on the luminosity of the individual star.  The most widely used luminosity indicator is a star's height, in magnitude space, above the horizontal branch, $V-V_{\rm{HB}}$.  This luminosity indicator is used because the relationship between $\Sigma$\,Ca and magnitude above the horizontal branch is very close to linear for iso-metallicity tracks.  Thus, by accounting for this linear relationship in magnitude space, the relation between \feh and $\Sigma$\,Ca becomes linear and can be used to directly infer the intrinsic metallicity of a star.  Due to the large distances of our dwarf galaxies, many do not have well determined horizontal branch (HB) magnitude.  The distance estimates, determined via the tip of the RGB, may have large errors for some of these galaxies, but this error is dwarfed by the error in $V_{\rm{HB}}$.  The error in $V_{\rm{HB}}$ may be as high as 0.5 magnitudes, resulting in a 0.1 dex error in the \feh measurement.   We instead utilize the absolute $I$-band magnitude, which has been shown to be more robust against age effects \citep{Carrera2007} and has an overall smaller error attached to its determination for our sample.

We determine \feh values for individual stars by combining the equivalent width and absolute I-band magnitude of RGB stars using a two-line calibration, to accommodate for the lower average S/N of our data, from the C13 dataset kindly provided by Ricardo Carrera (private communication):

\begin{eqnarray}
 \rm{[Fe/H]} & = & -3.51+0.12 \times {M_I} + 0.57 \times \Sigma \rm{Ca} \nonumber \\
 &&-\: 0.17 \times \Sigma \rm{Ca}^{-1.5} + 0.02 \times \Sigma \rm{Ca} \times {M_I}.
 \end{eqnarray}
 
 Where $\Sigma \rm{Ca}$ is the total equivalent widths of the measured Ca II lines and ${M_I}$ is the absolute $I$-band magnitude.  This calibration utilizes a sample of 422 stars in globular and open cluster and well as field stars to extend the CaT method to encompass the range from $-4.0 < \rm{[Fe/H]} < +0.50$ and over 5 orders of magnitude starting from the horizontal branch extending up to the tip of the red giant branch.

We determine the errors in $\Sigma \rm{Ca}$ by using 1000 Monte-Carlo re-samplings of each individual spectra with Gaussian random noise, scaled to the variance per pixel, added to each pixel.  The formal 1-$\sigma$ error is the square root of the variance about the mean of these 1000 realizations per star, for each of the two CaT lines.  We add in quadrature to these errors a systematic error of $0.25\mbox{\,\AA}$ calculated from repeat observations of 46 stars using the method outlined in \citet{Simon2007a,Willman2011}.  These errors are then appropriately propagated to the \feh calculations.

\subsection{Metallicity Sample Selection Criteria} \label{metal_samp}

\begin{figure}[!h]
\centering
\includegraphics[width=.48\textwidth]{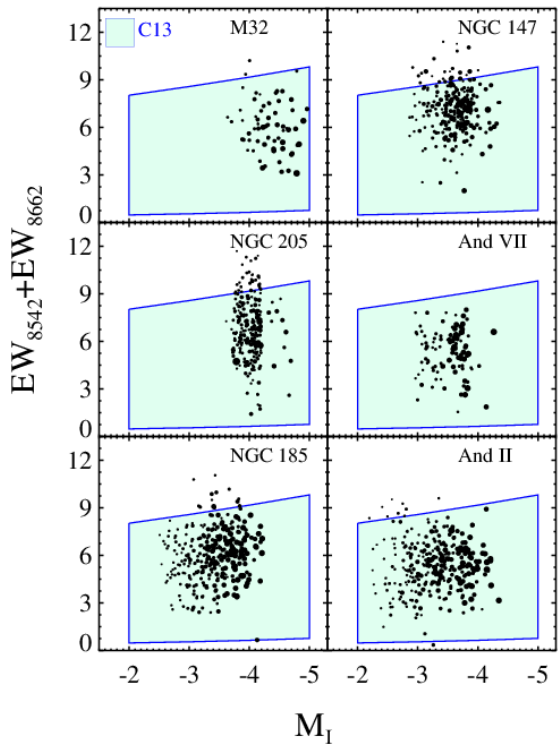}
\caption{Total CaT EW as a function of absolute $I$-band magnitude, ${M_I}$, for member stars with a measured calcium triplet metallicity.  Overlaid in light blue are the regions encompassed by the \citet{Carrera2013} calibration.  Symbol sizes correspond to the S/N of the individual star such that larger symbols represent stars with larger S/N and smaller symbols correspond to stars with lower S/N.} \label{EW_fig}
\end{figure}

Membership for the majority of our RGB stars were sourced from earlier publications (\S\,\ref{methods}).  These member samples were selected for kinematic work which has less stringent requirements on S/N and wavelength coverage than the present analysis.  We therefore impose additional criteria on the published samples.

\begin{figure}[!h]
\centering
\includegraphics[width=.35\textwidth,angle=90]{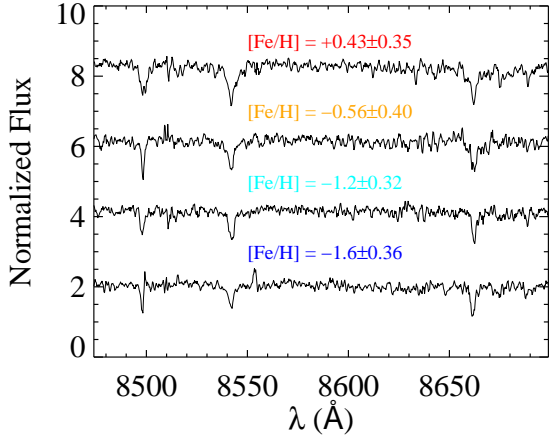}
\caption{Normalized, box-car smoothed spectra, with a smoothing window of two pixels, centered around the CaT region for four stars in NGC 147 with similar absolute I-band luminosities, ${M_I} \sim -$3.6.  These four stars have S/N values similar to the average S/N of all stars with a measured metallicity in NGC 147.  The top spectrum is that of a star with a CaT \feh = +0.43, the second spectrum with a \feh = $-$0.56, the third spectrum with \feh = $-$1.2, and the bottom spectrum with \feh = $-$1.6.  These spectra illustrate that, as metallicity decreases, the strength of the CaT lines at 8542 $\mbox{\,\AA}$ and 8662 $\mbox{\,\AA}$ decreases as well.} \label{SPEC_fig}
\end{figure}

We first remove member stars which do not have spectral coverage in the rest-frame continuum and line bandpasses of the CaT.  We next remove all stars with a continuum S/N$< 3$/pixel as the EW measurements are unreliable below this S/N threshold.  We motivate this S/N criteria by modeling our ability to recover the true EW by adding noise to a series of model spectra.  We found that, at a S/N $\sim 3$/pixel, the true EWs of the model spectrum could not be recovered within an error of 30\%.  To remove spectra with particularly bad sky subtraction, we impose a final criterium based on the reduced-$\chi^2$ fit of the Gaussian profile fit to the spectrum.  Of stars removed by the above cuts, 181 were removed due to incomplete spectral coverage, 478 by too low S/N and 174 due to poor sky subtraction in the CaT region.

In Figure~\ref{EW_fig}, we plot the region encompassed by the C13 calibration as function of the total equivalent width ($\Sigma $Ca) versus absolute \textit{I}-band magnitude.  We place our sample stars on this plot by converting the reddening corrected, apparent $I$-band magnitudes into absolute magnitudes using the distances listed in Table \ref{table_properties}.  Out of the original 2162 kinematic members, we measure a CaT EW for 1329 stars passing the above quality cuts.  Our final sample consists of 1243 stars which lie within the bounds of the calibration, shown in blue in Figure \ref{EW_fig}.  Stars outside of the calibration bounds tend to have larger errors on their EW measurements and represent a small portion of the overall sample.  For completeness, we include the C13 \feh values of stars outside the calibration window in subsequent figures, but do not include them in our average properties, MDF, or radial profile analysis.  In Figure~\ref{SPEC_fig} we show the spectra for four stars in NGC 147 with similar ${M_I}$ ranging from \feh = $+$0.47 to \feh = $-$1.6 and with S/N$\sim 6$/pixel, similar to the average S/N of the NGC 147 sample.  We provide in Table \ref{table_allstars} the derived \feh and associated error for the 1243 stars that have calibrated \feh.

Figure~\ref{CMD_fig} shows the CMD position of these metallicity-member stars (red) overlaid on all member stars (black).  For comparison, we show in gray the photometric sample that this spectroscopic data were sourced from.  Using previous photometric \feh values for each of these objects, we show bounding 12 Gyr and 2 Gyr Padova \citep{Girardi2002}  isochrones of \feh = $-$2.0, \feh = $-$1.0, and  \feh = $+$0.00.    

\begin{figure}[!h]
\centering
\includegraphics[width=.35\textwidth,angle=90]{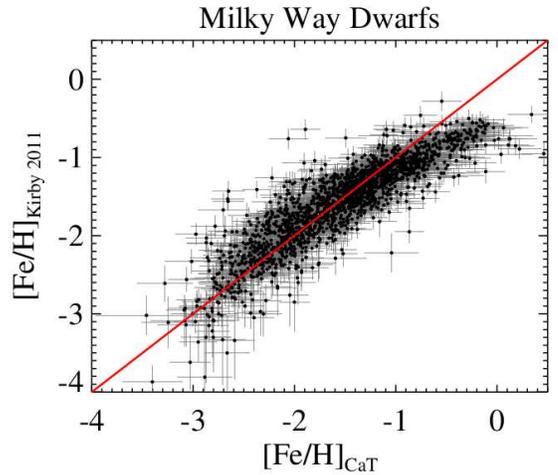}
\caption{Comparison between \feh values for 1,993 individual RGB stars in MW dSphs using the methods detailed in this paper and \feh values from K11's spectral synthesis method, along with requisite errors for each star.  The dwarfs with metallicities shown here are: Fornax, Leo~I, Sculptor, Leo~II, Sextans, Draco, Canes Venatici~I, and Ursa Minor.  The solid, red line represents the 1:1 relation.  At the low metallicity end, there is a 1:1 relation between \feh measured using the spectral synthesis technique and that using the CaT calibration from C13.  At \feh $>$ $-$1.0, the metallicities derived using the CaT begin to deviate from that of the spectra synthesis method.  This flattening at higher metallicities show that for the same stars, the CaT \feh derived from the C13 calibration is higher than that derived using the spectral synthesis technique.} \label{comparison_kirby}
\end{figure}

To confirm that our CaT \feh measurements using the C13 calibration are representative of the true \feh of the individual RGB stars, we apply the same analysis and selection criteria to the eight dwarfs from the K11 sample: Fornax, Leo I, Sculptor, Leo II, Sextans, Draco, Canes Venatici I, and Ursa Minor.   The K11 sample uses spectral synthesis to determine metallicities, a more robust method with significant more stringent S/N criteria.
We calculate \feh using our CaT method based on spectra kindly provided by E.~Kirby (private communication) and  compare to the published K11 \feh measurements, shown in Figure \ref{comparison_kirby}.  We find consistent \feh values for individual stars that are well within the measurement errors for all systems except Fornax.  We find a discrepancy in Fornax in that we find many more metal-rich stars (\feh $>$ $-0.5$) compared to that of K11.  The K11 spectral synthesis grid does not extend more metal-rich than \feh =0.0 and is calibrated to globular cluster stars only up to \feh $<$ $-0.5$.   This may be the source of discrepancy combined with possible differences in [$\alpha$/Fe] between open cluster RGBs and globular cluster RGBs used in the C13 calibration.  

We reproduce the \avgfeh, $\sigma([{\rm Fe/H}])$, and the shape of the MDFs, as well as radial trends reported for the K11 galaxies using the C13 CaT calibration.  In this paper, we highlight results from our CaT calibration only for the three most luminous systems from the K11 sample: Fornax, Leo I, and Sculptor because these have comparable luminosities to our six M31 dwarf galaxies.

\subsubsection{Biases in Metallicity Distributions}

To determine how representative our sample metallicities are of the true distribution of metallicity within each system, we investigate possible biases in our sampling.  Incomplete or biased sampling of the true underlying metallicity of a galaxy could skew the shape of the MDF, the radial profile, and the observed average properties.  These biases include: incomplete sampling along the RGB track, insufficient S/N for a metallicity sub-population, preferentially selecting a metallicity sub-population due to color cuts, or incomplete spatial coverage of the galaxy.    

For single-aged stellar populations, the metallicity of a star is correlated with its apparent magnitude such that more metal-rich stars reside in the lower reaches of the RGB.  Thus, due to observational limitations, metal-rich RGBs are usually also those with the lowest S/N compared to brighter, more metal-poor stars.  \citet{Reitzel2002} showed that by culling the stellar sample by magnitude, higher-metallicity stars were preferentially removed as they were generally fainter and lie father down in the RGB.  Using a magnitude cut-off based on where the luminosity function of member stars turn over, we split the sample in two and compare the resulting MDFs.  We find that stars below this threshold tend to contribute more metal-rich stars to the MDF than those above the turnover, however the metallicity range covered by each respective sample remains the same.  Similarly, to investigate the effects of S/N cuts, we divided the sample in half using a S/N such that there are approximately the same number of stars above and below this S/N sample bisector.  We then compared the \avgfeh and shape of the resulting MDF for the split sample and find no significant difference between their properties.  This lack of difference between the split samples is because stars of all S/N lie along all positions in magnitude and color space due to variations in integration times between different masks, thus we do not expect there to be a difference in metallicity space when cutting by S/N. 

The spatial coverage for our sample is excellent for And~II, And~VII, NGC~147, and NGC~185 all of which have coverage over the full radial extent of the galaxy.  For M32, the observations are limited to the outer edge of this galaxy ($> { 2r_{eff}}$) due to extremely high stellar density in the central region.  Similar to M32, NGC~205 lacks coverage in the innermost regions due to the high stellar density resulting in a lack of distinct sources.  However because it is not as compact as M32, we were able to probe within the effective radius.  In these two cases, spatial coverage exterior to these regions was excellent and well covers the observable area of the dwarfs.  For And~VII, our sample represents the most spatially complete sample thus far with coverage extending beyond two times the effective radius.  For all of our galaxies, there are no strong spatial biases in the metallicity results because the spectroscopic data encompass an area equal to that of the photometric sample for each dwarf in which there were discrete sources.  

Finally, our color selection for And~II, And~VII, NGC~147 and NGC~185 may introduce a metallicity bias into our sample.  Metal-rich stars tend to have redder colors than more metal-poor stars for a single-aged stellar population, thus any color selection on either the red-ward or blue-ward side could be removing a portion of either the metal-rich or metal-poor ends.  To determine this effect, we split our sample by the average metallicity in each system into "metal-rich" and "metal-poor" subsamples.  We then examined the CMD position of these two metallicity sub-samples and, for all galaxies except for NGC 185, do not find a substantial difference in their properties using a 2-D K-S test; all had p-values $>$ 0.01.  This shows that \feh does not correlate with color, consistent with our sample containing many aged stellar populations.  For NGC 185, despite differences in the color distribution of metal-rich and metal-poor subsamples, the broadness of our color selection minimizes any preferential selection of a metallicity population.  Given this, any bias that may have been introduced by color selection does not appear to have a large effect in our final metallicity distributions for the six dwarfs.  

\section{Results and Discussion}

\subsection{Global Metallicity Properties: Average \feh}\label{ssec_global_feh}

For our six M31 dwarf galaxies, we determined the average metallicities and intrinsic metallicity dispersion.  These are the first homogenous metallicity determinations of individual stars for three of our six M31 dwarf satellites: M32, NGC 205, and And VII.  Previous works on determining the global metallicity properties for M31 dwarf satellites have relied on photometric metallicity estimates from isochrone fitting to the observed CMD \citep[e.g.,][]{Kalirai2010} or coadded spectra of many stars \citep{Collins2013}.    Errors associated with these methods are large, ranging from 0.1 to 0.7 dex \citep{Collins2013}, compared to measuring stellar metallicities star by star and fail to encapsulate the internal metallicity dispersion and distribution of these dwarf systems.

\begin{figure}[!t]
\centering
\includegraphics[width=.7\textwidth,angle=-90]{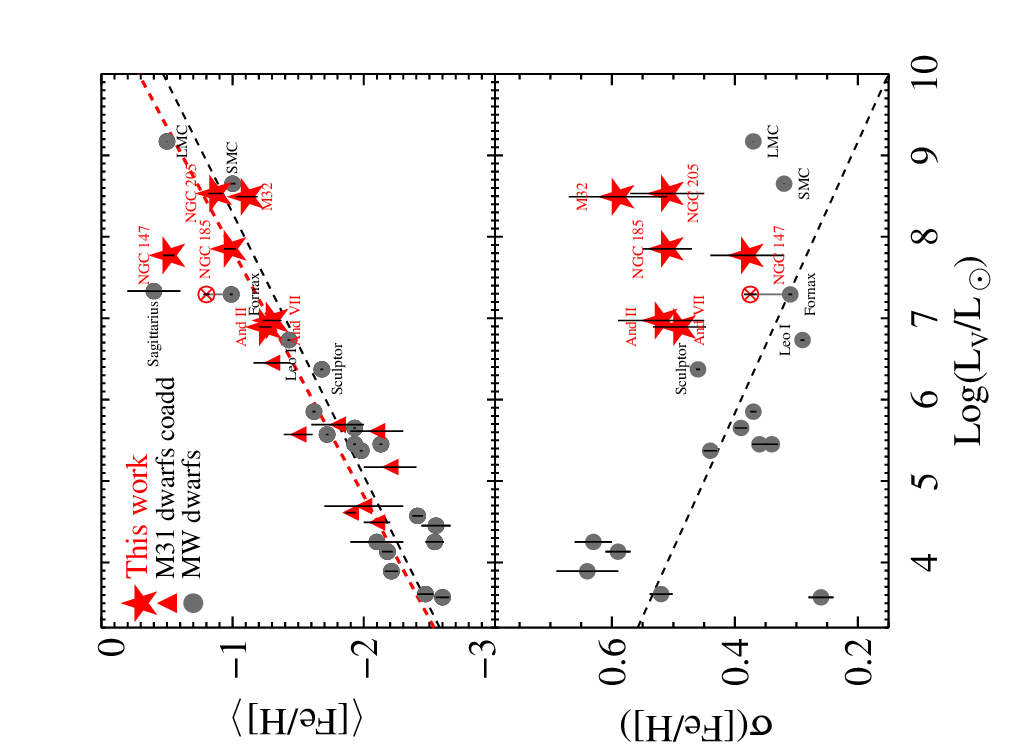}

\caption{\textit{Top}: Average metallicity of Milky Way (black symbols) and M31 (red symbols) dwarf galaxies as a function of luminosity.  Our six dwarfs are shown as red stars while red triangles are based on coadded spectra from \citet{Collins2013}.  This demonstrates that the M31 dwarf follow the same metallicity-luminosity relation as the MW dwarfs. \textit{Bottom}: Metallicity dispersions as a function of luminosity.  Contrary to the trend derived in K11 (dashed, black line), we do not observe an anti-correlation between intrinsic metallicity dispersion and luminosity. } \label{LUM_fig}
\end{figure}

We determine the average metallicity for each galaxy using a weighted sum such that the weight is inversely proportional to the errors on the metallicity measurement.  We find that for M32, the average metallicity of its outer regions is \avgfeh = $-1.11\pm0.09$.  This agrees with integrated spectroscopy metallicity measurements of \citep{Coelho2009} in the outer regions of M32; our observations do not cover the inner regions of M32 for which photometric metallicity estimates by \citet{Monachesi2011} suggest a more metal-rich (\avgfeh = $-0.20$) population.   For NGC 205, the average metallicity of the entire population, including tidal tails, is \avgfeh =  $-0.87\pm0.05$, consistent with the photometric metallicity estimates of \feh = $-0.8$ from \citet{Mcconnachie2005}.  Our CaT \avgfeh~measurements for NGC 147 and NGC 185 are more metal-rich at \avgfeh = $-0.51\pm0.04$ and \avgfeh = $-0.98\pm0.05$ respectively, compared to the \avgfeh = $-1.1\pm0.1$ and \avgfeh = $-1.3\pm0.1$ values found by \citet{Geha2010}.  This discrepancy is mainly due to differences in the calibration as \citet{Geha2010} used the Rutledge calibration, which is linear across the entire \feh regime whereas the C13 calibration is non-linear at both the metal-rich and metal-poor end.  The most recent \avgfeh measurement for And VII was done using KECK/LRIS by \citet{Grebel1999} with a reported \avgfeh = $-1.40$, consistent with our measured \avgfeh = $-1.30\pm$0.07.  Finally, in \citet{Ho2012} the metallicity for And II utilized the Starkenburg calibration and direct integration of the line bandpass regions; they found an \avgfeh = $-1.39\pm$0.03;  in this work we find an \avgfeh = $-1.25\pm$0.05, slightly more metal-rich than their findings.

As a group, our six dwarfs follow the metallicity-luminosity relation seen in the MW and M31 dwarfs \citep{Kirby2011,Mcconnachie2012,Collins2013}.  In Figure \ref{LUM_fig}, \textit{top panel}, we place our dwarfs (filled, red stars) on the metallicity-luminosity plane alongside all MW satellites with individual RGB metallicity measurements (filled, gray circles) and M31 satellites with coadded CaT metallicities (filled, red triangles).  Using the combined sample of M31 dwarfs with coadded or individual spectroscopic metallicity measurements, we derive the following relation between luminosity and metallicity:

\begin{eqnarray}
 \!\rm{\langle[Fe/H]\rangle} = (-1.94\!\pm\!0.09)\verb!+!(0.33\!\pm\!0.05)\rm{log}\!\left(\!\frac{L_{V}}{10^5L_{\odot}}\right)\!.
 \end{eqnarray}
 
The resulting relation using just M31 dwarfs with spectroscopic measurements (dashed, red line in Figure \ref{LUM_fig}, \textit{top})  is similar to the relation observed in K11 for just MW dwarfs (dashed line in Figure \ref{LUM_fig}, \textit{top}) and for Local Group dwarfs from \citet{Kirby13}.  The linear Pearson coefficient for this combined sample is 0.87, showing that luminosity and \avgfeh are highly correlated.  This positive trend between galaxy luminosity and galaxy \avgfeh has been shown to be linear for more massive galaxies, with a flattening for the most massive galaxies \citep{Tremonti2004, Andrews2013}.

\subsection{Global Metallicity Properties: Internal Metallicity Dispersion} \label{ssec_global_dispersion}

While in general mass appears to regulate the retention of metal-rich gas in a galaxy, its relation to the details of the star formation process is still nebulous.  To address this, we can look at the internal spread in metallicity which gives clues to the timescale of star formation.  K11 showed that the intrinsic spread in a galaxy's \feh is anti-correlated with its luminosity for the MW dwarfs with more massive systems having a smaller internal metallicity spread than lower luminosity systems.  

To determine whether our six M31 satellites obey this observed relation, we calculate the internal metallicity dispersion using the method described in K11.  In K11, the metallicity dispersion is defined to be the second moment about the mean of the distribution, which allows us to describe a dispersion that is distribution-independent.  We perform a Monte-Carlo bootstrap to determine the errors on the dispersion measurement using 1000 re-samplings drawn from a normal distribution that is scaled to the metallicity error for each individual star.  In order to fairly compare the samples,  we reproduce the average metallicities and internal metallicity dispersion, as well as place error bars on the metallicity dispersion measurements, for the fourteen galaxies from K11 used to construct this relation, as seen in Figure \ref{LUM_fig}, \textit{bottom}.  Additionally, given the discrepancy in the derived metallicities for Fornax stars, we include both the K11 metallicity dispersion,~\sigmafeh = 0.31, as well as that from our work using the C13 calibration, \sigmafeh = 0.42$\pm0.01$.

Adding to the results of \citet{Leaman2013}, who found that the anti-correlation observed in K11 flattens out higher luminosities, we find that the anti-correlation observed in K11 is completely removed for more luminous systems ($\gtrsim 10^5 L_{\odot}$), as shown in Figure \ref{LUM_fig}, \textit{bottom}. For the combined sample we find a linear Pearson coefficient of -0.18.  This shows a weak correlation between internal metallicity spread and luminosity, in contrast to K11. The lack of relation between  intrinsic metallicity dispersion and luminosity at  higher luminosities may be explained by the different ways in which star formation occurs in lower mass galaxies compared to more massive systems.  In lower mass galaxies, star formation is more stochastic due to these galaxies' inability to retain metal-enhanced gas from outflow events such as supernova winds.  For more massive systems, their deeper potential wells means that they are more resilient against wind-driven outflow events.  This ability to retain metal-enhanced gas means that star formation can proceed smoothly, compared to less massive systems.

\subsection{Metallicity Distributions}\label{ssec_mdfs}

\begin{figure*}[!t]
\centering
\includegraphics[width=.75\textwidth,angle=90]{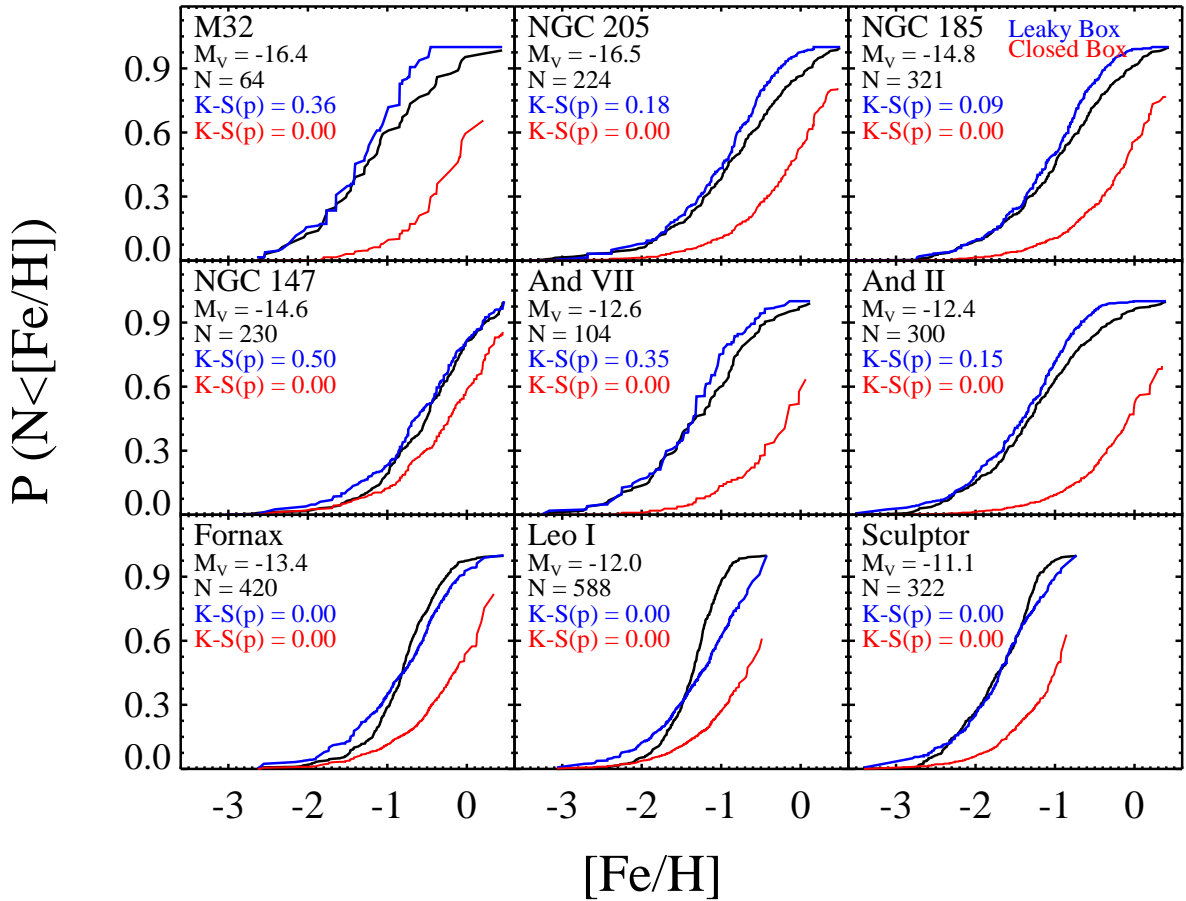}
\caption{Cumulative metallicity distribution functions (CMDFs) for our six dwarf galaxies (top two rows), along with three MW dwarfs (bottom row) with similar luminosities to And~VII and And~II, the two least luminous dwarfs in our sample.  Solid, black lines represent the CMDFs of all stars within the bounds of the C13 calibration.  The best-fit, Levenberg-Marquardt $\chi^2$ minimized, error-convolved leaky box (blue) models are presented.  Using a one-sided K-S, we determine the probability that our observed CMDFs are drawn from the leaky box chemical evolution models.  For all six of our M31 dwarfs, we find that the observed CMDFs of these dwarfs are consistent with being drawn from a leaky box distribution.  Our MW dwarfs comparison sample of Fornax, Leo~I, and Sculptor all have p-values that are inconsistent with being drawn from a leaky box distribution, in accord with the results from K11.  The resulting p-values for each of the nine dwarfs are labeled for the leaky box (blue).} \label{CMDF_fig}
\end{figure*}

\begin{figure*}[!t]
\centering
\includegraphics[width=.75\textwidth,angle=90]{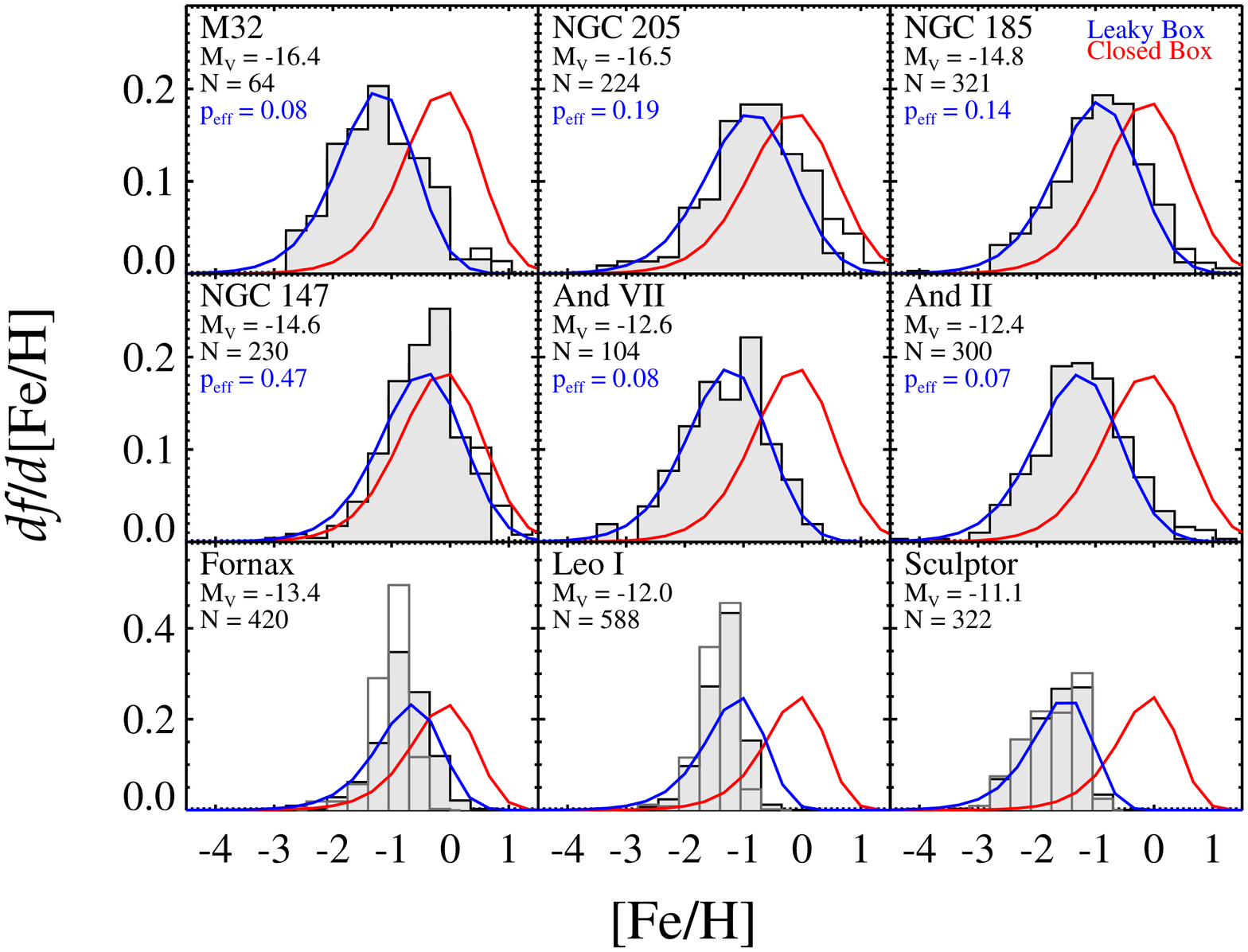}
\caption{Normalized metallicity distribution functions (MDFs) of our six galaxies (top two rows) along with three MW dwarfs (bottom row) with similar luminosities to And~VII and And~II, the two least luminous dwarfs in our sample.  In the top two rows black bordered, gray filled histogram represents the metallicity distribution functions of all stars within the bounds of the C13 calibration while unfilled portion shows those stars which lie outside the bounds of the calibration, but have metallicities calculated using the C13 calibration.  In the bottom row, gray-lined histograms represents the metallicity distribution functions from the spectral synthesis method of \citet{Kirby2011} while filled, gray histograms represent the CaT [Fe/H] measurements using the methods outlined in this paper.  Overlaid solid, red lines are the error-convolved fits to the MDFs using a closed box model of chemical evolution while solid, blue lines are the error-convolved fit to the MDFs using a leaky box model.  The average error in our [Fe/H] measurements are $\ge$ 0.4 dex, similar to the bin size of 0.35 dex.} \label{MDF_fig}
\end{figure*}

The cumulative and differential metallicity distribution function (CMDF and MDF, respectively) of a galaxy represents an admixture of its integrated star formation history and gas accretion/expulsion history.   By fitting the shape of the metallicity distribution with a model of galactic chemical evolution, we can broadly quantify the gas dynamics associated with the star formation history of the galaxy.  For all of the galaxies in our sample, this work represents the first spectroscopically determined CMDF and MDF of a large number of stars in each galaxy.  

We fit two simple chemical evolution models, the closed box and leaky box \citep{Tinsley1980}, to the CMDFs of our six dwarfs.  In the closed box model the system starts from an initial gas and is self-enriched through instantaneous recycling and mixing of metal-rich gas, a byproduct of the star formation cycle.  The total mass of the system remains constant in the closed box model with no gas inflow or outflow allowed.  The leaky box model has the same basic assumptions as the closed box model, but allows for the outflow of metal-enriched gas.  In this paper, we utilize the general, cumulative form:

\begin{eqnarray}
\rm{N_{\star}(<\rm{[Fe/H])}} = \rm{A} \left[1-\rm{exp}\left(\frac{10^{\rm{[Fe/H]_0}}-10^{\rm{[Fe/H]}}}{\it{p}}\right)\right]
\end{eqnarray}

where A is a normalization factor that accounts for the amount of metal-rich gas remaining in the system, [Fe/H]$_0$ is the initial metallicity of the gas, and $p$ is the stellar yield.  In this work, we use the initial condition of a pristine gas as our larger observational errors do not allow for a robust derivation of [Fe/H]$_0$.  For the closed box model, $p$ is the true yield of the system with a constant value of $p = 1.0 Z_{\odot}$ \citep{Fulbright2006}.  In the leaky box, gas that is metal-enhanced due to the star formation process is allowed to exit the system, thus driving down the yield $p$ and reducing it to an effective yield, $p_{eff}$, which is expressed as a fraction of the solar yield, $Z_{\odot}$.  
 
 We perform Levenberg-Marquardt $\chi^2$ minimization to find the best-fit closed and leaky box models to our spectroscopic CMDFs.  To determine whether our observed CMDFs are consistent with either the leaky box or closed box models, we utilize a one-sided K-S test to the best-fit CMDFs for each model.  However, given that the model CMDFs are best-fit models to the data, the standard probabilities associated the test statistic no longer applies as it is not distribution independent.  Using the methods outlined in \citet{Feigelson2012}, we determined the probability distribution for each galaxy and the best-fit closed and leaky box models by utilizing a parametric bootstrap.  For each galaxy, we construct cumulative distribution functions for both the model and the data.  We then perform 1000 bootstrap resamples of the best-fit model and convolved each model with individual measurement errors drawn from the error distribution of each galaxy.  Given that the best-fit model is error free, we construct the reference error-convolved best-fit model by doing a running median of the 1000 error-convolved, bootstrapped model distributions.  The K-S statistic, D = $\rm{max}\rm{|\rm{F_{n}(x)-F(x)}|}$ where F(x) is the model and F$_n$(x) is the bootstrap of the model, is calculated for each bootstrap resample.  The final, ordered distribution of D for the 1000 bootstrapped samples represents the probability distribution of the null hypothesis.    The resulting D statistic, D$^*$, between this reference model and the data is then used to tabulate a p-value.   In this work, the null hypothesis that the observed and model distributions are the same is rejected if p $\le 0.01$.
   
The resulting error-convolved, best-fit leaky box (blue) models are shown in Figure \ref{CMDF_fig} along with the K-S p-value for each model.  Similar to previous Milky Way studies, we find that the closed box model is a poor fit for all six dwarf galaxies in our sample.  The corresponding closed box K-S p-value for all nine dwarf systems are p=0.0.  In contrast, the CMDFs of all six dwarfs in our sample are consistent with being drawn from a pristine, leaky box CMDF.  Due to the larger observational errors of our sample, we are unable to robustly discern a difference between a pristine, leaky box and a pre-enriched, leaky box.  We include the CMDFs of Fornax, Leo I, and Sculptor along with their best-fit leaky and closed box models as a comparison sample.  Consistent with K11, we find that neither Fornax, Leo~I or Sculptor have distributions consistent with either a closed box, or leaky box. 

Despite the CMDFs showing consistency between the observed data and leaky box model, there are inconsistencies which become more apparent when observing the shapes of the differential metallicity distribution (MDF).  We show in Figure \ref{MDF_fig} the MDFs of our six dwarfs and the three comparison MW dwarfs along with the error-convolved best fit models for both the closed and leaky box models.  These MDFs show that, similar to the MDFs of MW dwarfs, there is diversity in the shapes of the distributions from peaky such as NGC 147 to very symmetric in the case of And~II.  In order to further quantify the shapes of the MDFs, we calculate both the skewness ($\gamma$) and excess kurtosis ($\kappa$), for each of the six galaxies, shown in Table \ref{table_properties}.  The skewness of a distribution gives insight into its symmetry, with negative values representing those distributions with longer metal-poor tails and positive values representing distributions with longer metal-rich tails.  The excess kurtosis, defined as the fourth moment minus 3, quantifies the peakiness of a distribution.  Positive values indicate distributions that are more sharply peaked than a standard gaussian whereas those with negative values of excess kurtosis are broader than a standard gaussian. In the following paragraphs we will discuss in detail the shapes of the MDFs for each dwarf and compare them to previous photometric works.

\textbf{M32:} The MDF presented in this paper is the first spectroscopic MDF composed of individual stars from M32.  Despite the sparse sampling limited to the outer regions of M32, the MDF is broadly consistent with previous works at the same radial distance \citep{Coelho2009}.  The width of the MDF is large with a span of almost three dex, comparable to previous photometric work which show similar broadness in the RGB population \citep{Grillmair1996, Monachesi2011}.  The leaky box provides a consistent fit with the CMDF of M32 with a best-fit effective yield of $\rm{p_{eff}}$ =  0.08.  This consistency shows that gas outflows have played a role in the chemical evolution of M32's outer regions.  

\textbf{NGC 205:}  The MDF of NGC 205 includes stars from both the main body of the galaxy as well as the tidal tails.  Similar to \citet{Butler2005}, we observe a long, metal-poor tail ($\gamma$ = $-$0.68) spanning over one dex in the MDF of this galaxy.  The bulk of the RGB stars in NGC 205, however, form a symmetric, fairly narrow peak ($\kappa$ = 0.52) around \avgfeh = $-$0.87 dex, similar to the position of the peak from the photometrically obtained MDF.  The closed-box model poorly fits the CMDF of this galaxy while the leaky box model provides a consistent fit with a best-fit effective yield of $\rm{p_{eff}}$ =  0.19.  However, upon further examination of the MDF, we can see that the number of metal-poor stars predicted by the leaky box model is higher than what is observed. 

\textbf{NGC 185:} The MDF of NGC 185 is asymmetric and has a thick, metal-poor tail ($\gamma = -$0.51) along with a somewhat broader peak region ($\kappa$ = 0.27) compared to NGC 205.  This broad peak was also observed photometrically by \citet{Butler2005}, who showed that this broadness is a result of an extended star formation history.  The best-fit leaky box model has an effective yield of $\rm{p_{eff}}$ =  0.14.  Similar to M32 and NGC 205, the leaky box model under-predicts number of metal-rich stars.  However, compared to NGC 205, the metal-rich side of the MDF of NGC 185 is much steeper.  This steepness on the metal-rich side could possibly be due to strong galactic winds which drive out metal-enhanced gas, preventing further metal-rich star formation \citep{Lanfranchi2004}.  

\textbf{NGC 147:} The MDF of NGC 147 is highly asymmetric ($\gamma = -$0.87) with a narrow peak region ($\kappa$ = 1.33), long metal-poor tail and a steep metal-rich cutoff.  This steepness on the metal-rich side was also seen in photometric MDFs from \citet{Butler2005}, however, our MDFs peaks 0.3 dex higher than that from photometric works.  The best-fit effective yield from the leaky box model is $\rm{p_{eff}}$ =  0.47, much higher than any other dwarf in the sample.  Accounting for the SFH from Geha et al.,\textit{in prep} in which the bulk of NGC 147 stars are of intermediate age may explain the higher \feh compared to the rest of the sample.  Furthermore, if this star formation proceeded rather quickly then the sharp peak in the MDF is not unexpected.  Another possible explanation for the higher \feh for NGC~147 compared to other galaxies at similar luminosity, is due to the current tidal interaction between M31 and NGC~147 \citep{Lewis2013,Ibata2013}.  Looking at the PAndAS maps from \citep{Lewis2013, Ibata2013} we see that the tidal tails for NGC 147 are very extended and span a projected size that is more than four times more extended than our spectroscopic data.  Additionally, these tails are comprised of mostly metal-poor stars \citep{Ibata2013} whereas the central regions of the galaxy are mostly metal-rich.  While we can not state how much mass is present in these extended tidal tails as that data has not been published, our observations of a higher than expected average \feh for NGC 147 can be explained by this tidal disturbance. 

\textbf{And VII:}  Very little work has been done on either the photometric or spectroscopic metallicity of And~VII other than to determine its \avgfeh.  The MDF of And VII is broad ($\kappa$ = $-$0.10) with an ill defined peak in its MDF and is more symmetric ($\gamma = -$0.41) than the NGC dwarfs of M31.  It's MDF is most similar to that of Canes Venatici I which was found by K11 to have an MDF consistent with being drawn from a leaky box model.  The leaky box model provides a consistent fit to both the metal-poor and metal-rich slopes of And~VII with a best-fit effective yield of $\rm{p_{eff}}$ =  0.08.

\textbf{And II:} The MDF of And II is fairly symmetric ($\gamma = -$0.15) with both a thick, peak region ($\kappa = -$0.21) and both metal-rich and metal-poor tails, consistent in shape to the photometric MDFs from \citet{Mcconnachie2007} and \citet{Kalirai2010}.  The leaky box model provides a good fit to both the metal-poor and metal-rich slopes with a best fit effective yield of $\rm{p_{eff}}$ =  0.07.  And II's star formation history appears to be extended with both old and intermediate aged populations in the horizontal branch \citep{Mcconnachie2007}.  This extended period of star formation could account for the broadness of the peak region while weaker galactic winds could account for the shallower metal-rich slope.
\\
\begin{figure*}[!t]
\centering
\includegraphics[width=.80\textwidth,angle=0]{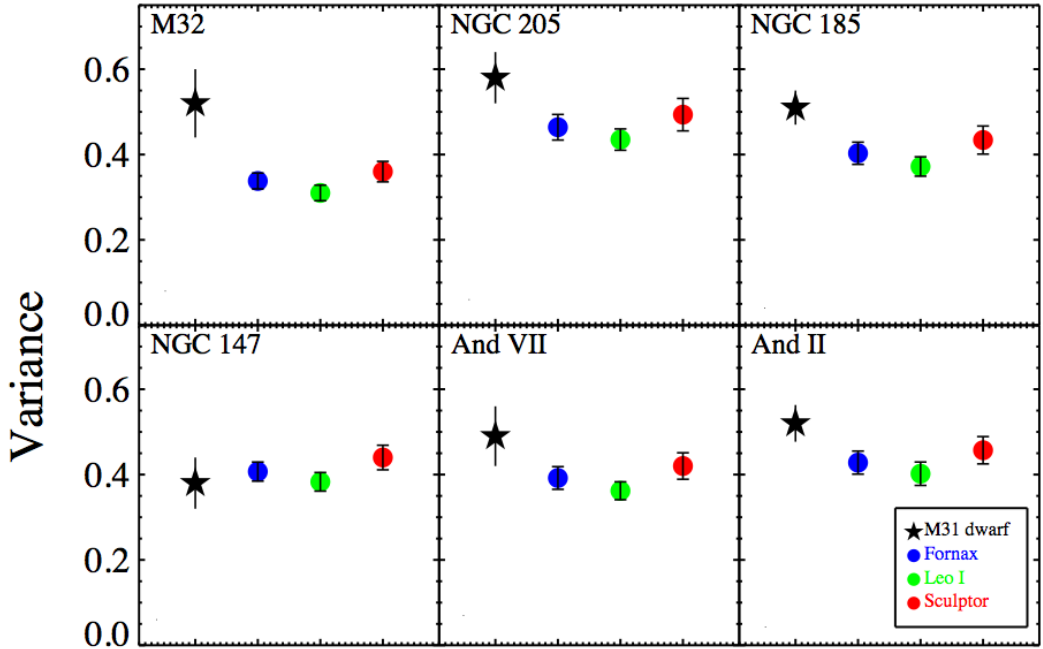}
\caption{In each of the six panels we show the second moment (variance) for each M31 dwarf (black star) along with the error-convolved variance of Fornax (blue circle), Leo~I (green circle), and Sculptor (red circle).  We use the error distribution associated with each reference M31 dwarf and convolved each MW dwarf with 1000 bootstrap resamples from this error distribution.  The values plotted here represent the mean-variance of each 1000 bootstrap resample along with $1-\sigma$ error bars calculated from the bootstrap.  We see that, even after convolving the MW dwarfs with the larger M31 errors, that for five out of our six galaxies their metallicity spreads are still larger than that of the MW comparison sample.  Thus, the broader MDFs observed in most of our dwarfs are an intrinsic property.} \label{Convolve_fig}
\end{figure*}

\begin{figure*}[!t]
\centering
\includegraphics[width=.7\textwidth,angle=90]{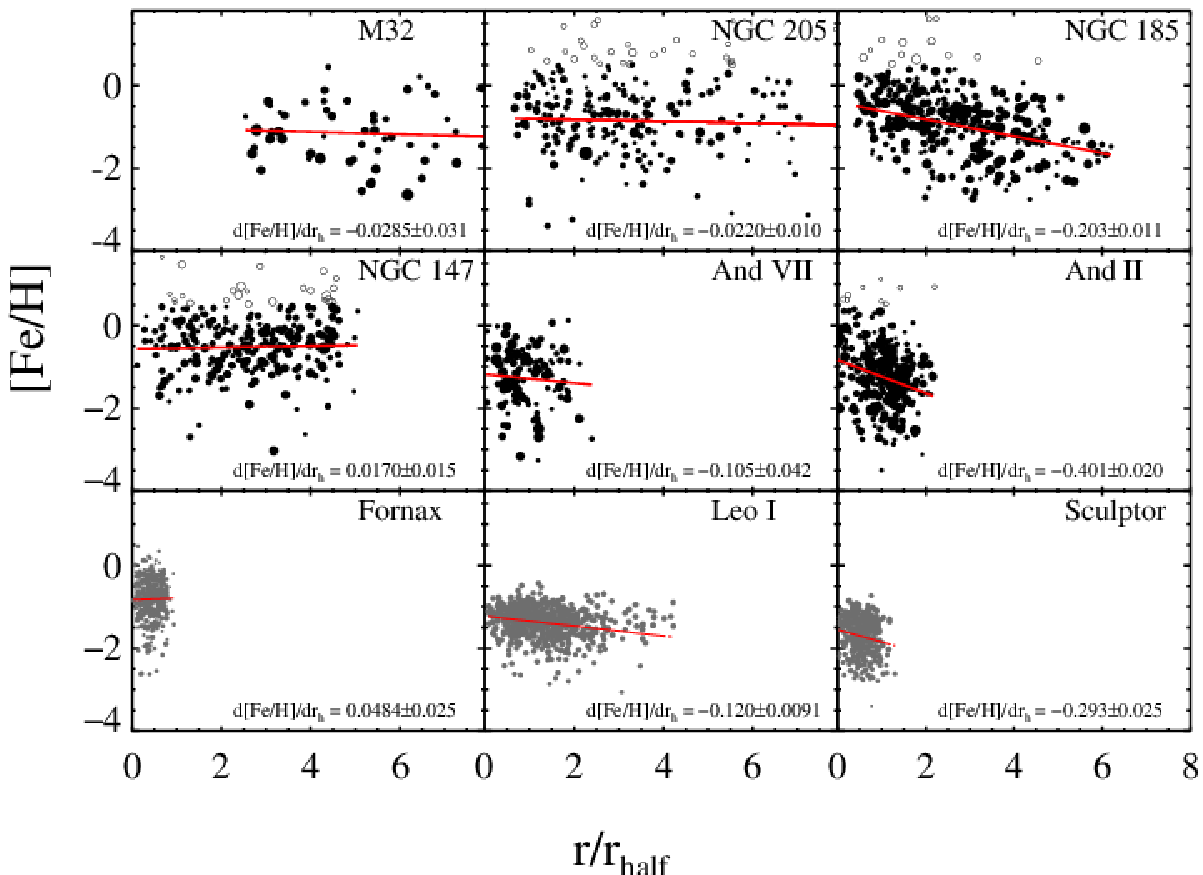}

\caption{Radial metallicity distribution of the six sample galaxies (top two rows) and three reference MW dwarfs (bottom row).  Black, filled circles are the M31 dwarf stars within the bounds of the C13 calibration while unfilled circles are stars outside of the calibration regions.  The individual symbol sizes are inversely correlated with their respective errors with larger symbols representing stars with smaller errors and smaller symbols representing stars with larger errors.  Overlaid in red, solid line is the best-fit line to the radial metallicity distribution of stars that lie within the bounds of the calibration (black, filled circles)} \label{Radial_fig}
\end{figure*}

Steepness on the metal-rich side of the shape of the MDF implies strong mass loss resulting in a sudden cessation of star formation.  The metal-rich gas that would have later formed into metal-rich stars was removed from the system via either environmental effects between the dwarf and its host halo environment or internal effects from the process of star formation.  Environmental effects, such as ram pressure or tidal stripping due to the interaction between a dwarf its host halo could remove gas from the system and halt further star formation \citep{GunnGott1972,FaberLin1983, PiatekPryor1995,Gnedin1999}.  Even if, individually these environmental effects are inefficient at gas removal, \citet{Mayer2006,Lokas2010} showed that a combination of ram pressure stripping and tidal shock heating is very efficient at gas removal.  Internal effects that could remove gas and inject energy into the dwarf's ISM are mainly wind driven, either by massive stars or supernovae \citep{Larson1974}.  The potential wells of these dwarfs are shallow enough that wind driven outflows could drive out metal-enhanced gas fairly quickly \citep{DekelSilk1986}.  However, HI observations of Local Group dwarfs by \citet{Grcevich2009} have shown that the gas-content of a dwarf is more correlated with its distance to a host halo than to its total mass or stellar content.  Thus, internal effects likely play a smaller role in halting star formation in local dwarfs.  

For most of our dwarfs, the MDFs show broad distributions that are more symmetric than MW dwarfs.  To determine if this broadness is an intrinsic effect or is due to our larger observational error, we convolve stars in the three MW dwarfs of comparable luminosity: Fornax, Leo~I, and Sculptor with the observational errors of stars from each dwarf galaxy in our sample.  We remove the observational error associated with each MW star by subtracting it in quadrature to a bootstrapped observational error sampled from the error distribution of each reference M31 dwarf.  We then computed the second moment (hereafter, variance), used to quantify the spread of the distribution, of these 1000 bootstraps.  Using the mean value this distribution of second moments as the value which quantifies the error-convolved spread for each MW dwarf, we compare it to the reference M31 dwarf.  We perform this method on all six dwarfs in our sample, producing a unique error convolved distribution of the three MW dwarfs for each M31 dwarf.  

The variance for each convolved MW dwarf are shown in Figure \ref{Convolve_fig} for reference.  For Leo~I (green), in no cases does the error-convolved variance become broader than the distribution of the reference M31 dwarf.  In contrast,Sculptor, which is known to have an MDF with large span due to an extended star formation history, has error-convolved MDFs with comparable spread to the M31 dwarfs.  For distributions which are intrinsically peaky, such as Leo~I, the inclusion of larger observational errors will broaden out the distribution, but the intrinsic shape is mostly preserved.  Likewise, distributions which are naturally broad such as Sculptor will only get broader, but again the intrinsic shape remains the same.  For Fornax (blue in Figure \ref{Convolve_fig}), which is more luminous than And~VII and And~II, the spatial sampling of our data includes only the central regions ($~1$ core radius).  \citet{Battaglia2006, Leaman2013}, both used a much larger sample that extends out to $~5$ core radii, have shown that Fornax possesses a metallicity gradient, which we do not observe using our data.  Thus, the variance value that we calculate is an under-estimate of the true variance of the Fornax distribution.  Using the variance calculation from \citet{Leaman2013}, variance = $~0. 47$, brings Fornax to consistency with our sample.  

While we expect larger observational errors to at least be partly responsible for the extra broadness observed in the M31 dwarfs, we do not expect them to be the sole contributors to this observed broadness.  Given that our dwarfs are in generally more massive than most MW dwarfs with derived MDFs, we expect them to be broader in general.  More massive systems are more resistant against metal removal events such as galactic winds and thus, are able to retain their gas and continue with star formation.  Less massive systems, due to their shallower potential wells, are more prone to both internally induced gas removal as well as externally induced gas removal such as tidal stripping.  This higher efficiency of gas removal means that the star formation histories of less massive systems are more easily truncated.

\subsection{Radial Metallicity Profiles}\label{ssec_radial}

In the Local Group, radial metallicity gradients are observed both photometrically and spectroscopically in many dwarfs such as Sculptor \citep{Battaglia2006}, Sextans \citep{Battaglia2011}, Fornax \citep{Tolstoy2004}, and Leo I \citep{Gullieuszik2009}, Leo II \citep{Koch2007}, and Draco \citep{Kirby2011}.  The presence, or absence, of a metallicity gradient are clues to the star formation, chemical enrichment, and dynamical history of these objects.  The details of the star formation process such as environmentally induced truncation or extension of star formation via gas expulsion or accretion can also be deduced based on the radial metallicity behavior.
  
Examinations of radial metallicity trends for some of our galaxies (M32: \citealt{Rose2005}, NGC 205: \citealt{Koleva2011}, NGC 147 and NGC 185: \citealt{Geha2010}, And II: \citealt{Ho2012}, and And VII: \citealt{Grebel1999}) have been previously presented based on 
either binned profiles or long-slit spectroscopy.   In Figure ~\ref{Radial_fig}, we show the radial metallicity trends for each galaxy from their constituent RGB population.  We plot distance from the galaxy center in units of elliptical half-light radii from \citet{Mcconnachie2012}.  While previous works such as K11 have used the core radius as a size indicator, here we use the half-light radius because the surface-brightness profiles of our dwarfs are poorly fit by King profiles. The red lines in Figure~\ref{Radial_fig} are the best-fit, error-weighted, linear least squares fit to the data.  We also determine radial metallicity distributions for Fornax, Leo I, and Sculptor as a function of core radius and compared them to the \citet{Kirby2011} distributions as a sanity check.  We find excellent agreement between the K11 results and our measured radial slopes as a function of core radius.  For uniformity, however, we show the radial metallicity profiles for these MW dwarfs as a function of half-light radius in the \textit{bottom row} of Figure \ref{Radial_fig}.

As shown in Figure~\ref{Radial_fig}, the radial metallicity distribution for our six dwarfs show the same diversity as the MW satellites.  Two of our dwarfs, NGC 185 and And II show significant radial gradients with stars in the outer radii being more metal-poor than those residing in the more central radii.  The slopes of M32, NGC 205, NGC 147, and And VII are all consistent with zero or nearly zero.  In no cases do we see a significant rise and subsequent flattening in the radial gradients in the outskirts, as was seen for Sculptor \citep{Battaglia2008}.  These trends do not change significantly with the addition of stars that fall outside the C13 calibration region.  In the case of NGC 185, including these stars increases the steepness of the radial gradient while for And II, the presence of two stars in the outer regions with higher metallicities slightly decreases the gradient slope.  However, because these stars are not anchored by the calibration, we do not include them in the analysis of the radial gradient slope.
 
Comparing our observed radial trends to previous works for each individual galaxy, we find that for four of our galaxies the trends are consistent with previous published works.  For NGC 205, the radial coverage in \citet{Koleva2011} extends out to the effective radius, whereas our sample extends out to well over five effective radii, but do not cover the innermost regions due to crowding issues.  Thus, while they see a radial trend in the inner regions of the galaxy, the difference in spatial coverage between our sample and theirs allows for the validity of both results.  However, for NGC 185 our results are inconsistent with that of \citet{Geha2010}, who observed a flat radial trend for NGC 185.  The difference between our work and theirs is two-fold: we utilize a different metallicity calibration and our profiles are unbinned, which avoids smoothing out the signatures of any possible gradient.  

For our sample of six M31 dwarfs along with the three MW dwarfs, we do not observe trends between slope of the radial gradient as a function of luminosity, distance from host, or dynamical state.  Four of our dwarfs, NGC 205, NGC 185, NGC 147, and And II are rotationally supported, yet only NGC 185 and And II possess radial gradients.  NGC 205 and NGC 147 are flat across five half-light radii, showing that radial coverage is not an issue and that, even at large radii, the metallicity spread is similar to the inner radii.  This lack of a trend between rotational support and the presence of a radial gradient is in tension with the observations of \citet{Leaman2013}, who found that $v/\sigma$ and the slope of the gradient are correlated for their sample which includes the LMC, SMC, WLM, Fornax, and Sculptor.  If we put our dwarfs on this relation, we find that our sample introduce very large scatter in the relation such that it is no longer statistically significant.

The radial metallicity behavior of each galaxy holds clues on the dispersion of metals during the star formation process.  Flat radial gradients can be a product of either a well mixed interstellar medium or from tidal interactions between a dwarf and its parent galaxy.  Galaxy wide mixing of gas could be due to wind driven gas expulsion of gas; if a galaxy is massive enough this gas eventually is recaptured and rains back onto the system \citep{Maclow1999}.  Tidal interactions between a dwarf and its parent galaxy can both disrupt stellar orbits and drive the inflow of gas into the system, resulting in additional pockets of star formation \citep{Mayer2006}.  NGC 205 and NGC 147 have both been shown to be undergoing tidal disruption by M31 \citep{Choi2002,Lewis2013,Ibata2013}, which may explain their flat gradients.   

Strong radial gradients, on the other hand, could be a result of purely internal evolution or external inflow of metal-rich gas.  Internal evolution leading to a strong radial gradient requires that the galaxy be kinematically undisturbed and that the star formation proceed over a longer span such that metals expelled from previous generations of stellar evolution have time to gravitationally sink toward the center of the galaxy \citep{Schroyen2013}.  This scenario may explain NGC~185 and And~II's strong radial gradient. NGC~185's star formation history has been shown by \citet{Butler2005} to be quite extended.  \citet{Mcconnachie2007} showed that And~II possesses an intermediate aged population that is centrally concentrated compared to the older stellar population, the superposition of two stellar populations with difference concentration indices may explain the strong gradient observed. External mechanisms leading to a metallicity gradient include inflow of metal-rich gas, however this requires the gas to be funneled directly into the central regions of the galaxy and setting off star formation.  However, in order to better determine which of these mechanisms are responsible for the radial trends in these galaxies, accurate ages and stronger constraints on the star formation histories of these objects are required.     

\section{Conclusion}

In this paper we present detailed metallicities of individual RGB stars within the six most luminous M31 dwarf satellites.  Using a new CaT metallicity calibration, we obtained \feh values from direct measurements of the CaT lines in individual RGBs.  We presented the global properties, metallicity distributions, and radial metallicity profiles for these six dwarfs.  This sample represents the first homogeneous spectroscopic metallicity analysis of individual stars within these galaxies and, as a whole, the most complete work on metallicity for the M31 dwarf system.  We summarize our results as follows:

\begin{itemize}

\item The \avgfeh of these six dwarfs places them on the observed [Fe/H]-luminosity relation of MW and M31 dwarfs.  Contrary to K11, we do not observe a negative trend between \sigmafeh and luminosity such that more luminous galaxies have smaller $\sigma$([Fe/H]).  Instead, we observe that the trend disappears with the inclusion of more luminous dwarfs.  However, without available data for \sigmafeh that cover the full spatial extent of the LMC, SMC, and Sagittarius, we can not discard the possibility that this lack of a trend is only observed for luminous M31 dwarfs.  Future work on the metallicity of the newly discovered luminous M31 satellites \citep{Martin2013}, as well as detailed metallicity studies of fainter M31 satellites, should answer whether the metallicity dispersion trends seen in M31 is representative of all dwarf galaxies, or only that of M31 satellites.  

\item The observed MDFs for our dwarfs are in general broader than that observed for MW dwarfs, most of which are less luminous than the dwarfs in our sample.  Even after accounting for larger observational error inflating the MDF of our galaxies, we still observe this enhanced broadness.  The agreement between the leaky box model and our observed metallicity distributions show that gas outflows played a role in the star formation histories of these galaxies.  However, due to the large observational errors of our sample, we are unable to discern the importance and strength of additional mechanisms such as outflow or inflow rate in guiding the chemical evolution of these galaxies.  

\item The radial metallicity profiles of these six dwarfs show the same diversity that has been observed for MW dSphs from K11 and for other local dEs \citep{Koleva2011}.  We do not observe any trend between galaxy luminosity, distance from host, or dynamical state and the presence of a radial gradient.  Of our six dwarfs, only two: NGC 185 and And II show evidence of radial metallicity gradients.  In order to determine the physical mechanisms required to produce these trends, accurate stellar ages and star formation histories are required, which we leave for future works.   

\end{itemize}

\section{Acknowledgements}

We would like to thank the anonymous referee for helpful suggestions that improved this paper.  MG acknowledges support from NSF grant AST-0908752.  EJT acknowledges that the support for this work was provided by NASA through Hubble Fellowship grant 51316.01 awarded by the Space Telescope Science Institute, which is operated by the Association of Universities for Research in Astronomy, Inc., for NASA, under contract NAS 5-26555.  

\bibliography{bib_mdf}
\bibliographystyle{apj}

\begin{deluxetable*}{ccllllrrcll}[!h]
\tabletypesize{\scriptsize}
\tablecaption{Summary of M31 Dwarf Metallicity Properties}
\tablewidth{0pt}
\tablehead{
\colhead{Dwarf} &
\colhead{Distance}&
\colhead{M$_V$} &
\colhead{N$_{\rm star}$}&
\colhead{\avgfeh} &
\colhead{$\sigma_{[Fe/H]}$} &
\colhead{Skewness}&
\colhead{Kurtosis}&
\colhead{R$_{\rm{half}}$} &
\colhead{$d\rm{[Fe/H]}/d(r/r_{h})$} &
\colhead{p$_{\rm{eff}}$}\\
\colhead{}&
\colhead{kpc}&
\colhead{} &
\colhead{} &
\colhead{dex} &
\colhead{dex} &
\colhead{}&
\colhead{}&
\colhead{pc}&
\colhead{dex }&
\colhead{Z$_{\odot}$}
}
\startdata
M32 &  805 & $-$16.4 & 64 & $-$1.11$\pm$0.08 & 0.59$\pm$0.08 &0.003$\pm$0.06& $-$0.70$\pm$0.17 & 110 & $-$0.03$\pm$0.03 & 0.08$\pm$0.003 \\
NGC 205 &  824 & $-$16.5 & 224 & $-$0.87$\pm$0.05 & 0.51$\pm$0.06 &$-$0.68$\pm$0.03& 0.52$\pm$0.15  & 590 & $-$0.02$\pm$0.01 & 0.19$\pm$0.004 \\
NGC 185 & 676 & $-$14.8 & 321 &  $-$0.98$\pm$0.05 & 0.51$\pm$0.04 &$-$0.51$\pm$0.04& 0.27$\pm$0.41  & 623 & $-$0.20$\pm$0.01 & 0.14$\pm$0.002\\
NGC 147 & 617 & $-$14.6 & 230 & $-$0.51$\pm$0.04 & 0.38$\pm$0.06 &$-$0.87$\pm$0.04& 1.33$\pm$0.28  & 458 & $+$0.02$\pm$0.02 & 0.47$\pm$0.005\\
And VII & 762 & $-$12.6 & 104 & $-$1.30$\pm$0.07 & 0.52$\pm$0.07 &$-$0.41$\pm$0.06& $-$0.10$\pm$0.30  & 776 & $-$0.11$\pm$0.04 & 0.08$\pm$0.002\\
And II & 652 & $-$12.4 & 300 & $-$1.25$\pm$0.05 & 0.49$\pm$0.04 &$-$0.15$\pm$0.02& $-$0.21$\pm$0.14  & 1176 & $-$0.40$\pm$0.02 & 0.07$\pm$0.001
\enddata
\tablecomments{The columns are ordered as follows: (1) Galaxy name,  (2) Distance to the galaxy, (3) Absolute V-band magnitude from , (4) number of stars in the final metallicity sample, (5) average, weighted mean metallicity with associated errors, (6) internal metallicity dispersion, calculated by taking the second moment of the distribution and accounting for increased dispersion from observational errors, along with Monte-Carlo errors, (7) the skew, or third moment, of the distribution with associated errors, (8) the excess kurtosis, or fourth moment minus 3, of the distribution with errors, (9) half-light radius, (10) the radial metallicity gradient as function of half-light radius with associated errors from a least squares fit, and (11) the best-fit value of the stellar yield resulting from fitting the leaky box model to the MDF of each dwarf.  Values from columns (2), (3), and (9) were taken from \citet{Mcconnachie2012}.} \label{table_properties}
\end{deluxetable*} 

\begin{deluxetable*}{cccccccc}[!h]
\tabletypesize{\scriptsize}
\tablecaption{Metallicity of individual stars in each galaxy}
\tablewidth{0pt}
\tablehead{
\colhead{Star ID} &
\colhead{Galaxy Name}&
\colhead{RA} &
\colhead{DEC}&
\colhead{V-I} &
\colhead{V} &
\colhead{\feh} &
\colhead{[Fe/H]$_{\rm{error}}$} \\
\colhead{}&
\colhead{}&
\colhead{(h$\,$:$\,$m$\,$:$\,$s)} &
\colhead{($^\circ\,$:$\,'\,$:$\,''$)} &
\colhead{} &
\colhead{} &
\colhead{dex}&
\colhead{dex }
}
\startdata

1&M32&0:42:48.50&40:47:24.20&1.6&21.36&$-$1.16&0.36\\
2&M32&0:42:45.10&40:48:05.40&1.6&21.64&$-$0.84&0.36\\
3&M32&0:42:52.30&40:48:53.20&2.2&22.08&$-$0.06&0.41\\
4&M32&0:42:47.05&40:49:34.40&1.5&21.66&$-$0.84&0.39\\
5&M32&0:42:41.79&40:49:36.80&2.0&21.81&$-$2.55&0.37\\
6&M32&0:42:52.50&40:49:37.50&2.1&22.48&$-$1.45&0.48\\
7&M32&0:42:56.27&40:49:38.10&2.6&22.69&$-$0.45&0.42\\
8&M32&0:42:41.63&40:49:46.00&1.8&21.64&$-$0.37&0.38\\
9&M32&0:42:38.25&40:50:56.60&1.3&21.54&$-$1.13&0.38\\
10&M32&0:42:51.25&40:51:07.90&3.6&23.34&$-$1.84&0.35\\
11&M32&0:42:47.46&40:51:09.70&1.6&21.15&$-$1.28&0.32\\
12&M32&0:42:44.16&40:52:56.60&1.9&21.38&$-$1.07&0.27\\
13&M32&0:42:34.43&40:53:04.70&1.7&21.19&$-$1.76&0.28\\
14&M32&0:42:46.72&40:54:13.60&1.4&21.15&$-$0.09&0.35\\
15&M32&0:42:46.62&40:48:01.50&2.4&22.63&$-$1.52&0.42\\
16&M32&0:42:52.74&40:50:41.90&1.9&22.24&$-$1.24&0.38\\
17&M32&0:42:34.13&40:50:43.70&1.8&21.84&$-$1.08&0.39\\
18&M32&0:42:47.34&40:52:55.40&1.4&21.74&$-$1.76&0.56\\
19&M32&0:42:40.30&40:53:52.80&1.6&21.36&$-$1.23&0.62\\
20&M32&0:42:45.63&40:54:24.00&1.8&21.82&$-$1.43&0.46\\
21&M32&0:42:51.69&40:48:37.80&1.9&21.96&$-$1.05&0.40\\
22&M32&0:42:43.88&40:49:26.20&1.8&21.68&$-$1.09&0.33\\
23&M32&0:42:42.70&40:50:23.00&1.9&21.63&$-$1.28&0.31\\
24&M32&0:42:35.25&40:50:34.70&2.4&21.97&$-$1.78&0.43\\
25&M32&0:42:44.80&40:50:45.70&1.7&21.60&$-$1.50&0.44\\
.. & .. &..&..&..&..&..&..\\
.. & .. &..&..&..&..&..&..
\enddata
\tablecomments{The columns are ordered as follows: (1) Star ID,  (2) Host Galaxy, (3) Right Ascension, (4) Declination, (5) color, (6) extinction corrected apparent V-band magnitude, (7) metallicity, and (8) metallicity error.  The full table can be accessed from a machine readable table from the journal website.} \label{table_allstars}
\end{deluxetable*}

\end{document}